\documentclass[aps,fleqn,twocolumn]{revtex4}

\usepackage{graphicx}
\usepackage[mathscr]{eucal}
\usepackage{amsmath}
\usepackage{amssymb}
\usepackage{epstopdf}
\usepackage{epsfig}

\newcommand{\beq}{\begin{equation}}
\newcommand{\eeq}{\end{equation}}
\newcommand{\beqa}{\begin{eqnarray}}
\newcommand{\eeqa}{\end{eqnarray}}
\newcommand{\ket} [1] {\vert #1 \rangle}

\newcommand{\braket}[2]{\langle #1 | #2 \rangle}

\newcommand{\mean}[1]{\langle #1 \rangle}

\newcommand{\tr}{\mathop{\mathrm{tr}}}

\def\one{\ensuremath{\hbox{$\mathrm I$\kern-.6em$\mathrm 1$}}}
\def\tr{ \mbox{tr}}

\begin{document} 

\title{Scaling of entanglement support for Matrix Product States}

\author{L. Tagliacozzo$^1$, Thiago. R. de Oliveira$^{1,2}$, S. Iblisdir$^1$ and  J. I. Latorre$^1$}
\affiliation{$^1$ Dept. Estructura i Constituents de la Mat\`eria, Universitat de Barcelona, 08028 Barcelona, Spain \\
$^2$ Instituto de F\'isica Gleb Wataghin, Universidade Estadual de Campinas, CP 6165, CEP 13083-970, Campinas, SP, Brazil}

\begin{abstract}

The power of matrix product states to describe infinite-size translational-invariant 
critical spin chains is investigated. At criticality, the accuracy with which they describe
ground state properties of a system is limited by the size $\chi$ of the matrices that form
the approximation. This limitation is quantified in terms of the scaling of the half-chain 
entanglement entropy. In the case of the quantum Ising model, we find $S \sim \frac{1}{6}\log \chi$ with 
high precision. This result can be understood as the emergence of an effective
finite correlation length $\xi_\chi$ ruling all the scaling properties in the system.  We produce six extra pieces of evidence for this finite-$\chi$ scaling, namely, the scaling of the correlation length, the scaling of magnetization, the shift of the critical point, the scaling of the entanglement entropy for a finite block of spins, the existence of scaling functions and the agreement with analogous classical results. All our computations are consistent with a scaling relation of the form $\xi_\chi\sim \chi^{ \kappa}$, with $\kappa=2$ for the Ising model. In the case of the Heisenberg model, we find similar results with
the value $\kappa\sim 1.37$. We also show how finite-$\chi$ scaling allows to extract critical exponents. These results are obtained using the infinite time evolved block decimation algorithm which works
in the thermodynamical limit and are verified to agree with density matrix renormalization
group results and their classical analogue obtained with the corner transfer matrix renormalization group.
\end{abstract}

\maketitle

\section{Introduction}
%%%%%%%%%%%%%%%%%%%%%%

The exact solution of the dynamics of quantum physical systems is often 
too hard or impossible to compute. It is then necessary to resort to approximation
schemes and numerical simulations, as in the case of 
QCD, the theory of strong interactions, to gain some insight
into the physics of the theory under study. These numerical simulations
are implemented using some clever algorithm that exploits the 
understanding of the quantum interactions at work. It may then be
difficult to separate what is the absolute limitation inherent to the 
nature of the approximation from what is an artifact of the specific 
algorithm employed. 

We can elaborate further this idea in the case of one-dimensional
translational invariant spin chains. There, the recent algorithms  introduced by Vidal based
on the explicit use of Schmidt decompositions \cite{Vidal:first}
have been shown to deliver identical results to the
very successful Density Matrix Renormalization
Group (DMRG)\cite{White,Schollwock,hallberg}.
Actually, these two apparently wide-apart algorithms
agree because they 
come down to represent the coefficients of a quantum state as
a product of matrices, that is a Matrix Product State (MPS) 
\cite{ostlund,mpsrepr,mcculloch}
\beq
\ket{\Psi}=\sum_{s_1 \ldots s_N} \tr [A_1(s_1) \ldots A_N(s_N)] \ket{s_1 \ldots s_N},
\eeq
where $s_i$ labels a basis for the local degree of freedom ('spin')
 of particle $i$, and where the $A_i(s_i)$'s are matrices 
of some fixed finite size, $\chi$, and $N$ is the number of 
sites in the chain which will be taken to be infinite 
\footnote{Note that $\chi$ is often referred as $m$ in the context of DMRG.}. 
Under the assumption that the above mentioned algorithms 
do find a faithful description of the sought state, consistent with the MPS structure,
we can forget about their details and describe their results
as a consequence of the properties of MPS states.

In this paper, we shall investigate what is the limitation
attached to the use of the MPS approximation for infinite
one-dimensional translational invariant quantum systems. 
It is important to 
note that we address infinite systems in order to avoid the presence
of any finite size effect. 
Consequently, any departure of MPS results from the exact ones is 
expected to be due to the
very nature of the MPS representation and must necessarily
relate to the finite matrix-size $\chi$ that can be handled in practice.

Let us now present a brief summary of our main result. We need first to
recall the basic construction of the Schmidt decomposition for
any state in a bipartite Hilbert space ${\mathcal H={\mathcal H}_A\otimes
{\mathcal H}_B}$,
\beq\label{eq:schmidt}
\ket{\Psi}=\sum_{\alpha=1}^{\min(dim {\mathcal H}_A,{\mathcal H}_B)} 
\lambda_{\alpha} \ket{\varphi_{\alpha}^A} \ket{\varphi_{\alpha}^B},
\eeq
where $\sum_{\alpha} |\lambda_{\alpha}|^2=1$, 
$\braket{\varphi_{\alpha}^A}{\varphi_{\beta}^A}=\braket{\varphi_{\alpha}^B}{\varphi_{\beta}^B}=\delta_{\alpha \beta}$. 
The amount of entanglement  (quantum correlations) between $A$ and $B$ can be quantified in terms of the von Neumann entropy of part $A$ (or $B$):
\beq\label{eq:vnentropy}
S(\rho_A)=-\sum_{\alpha} \lambda^2_{\alpha} \log \lambda^2_{\alpha}=S(\rho_B)\ .
\eeq
This entanglement entropy
in an infinite chain is known to obey scaling properties. 
At a critical point \footnote{Throughout this paper, we will only consider second order phase transitions.}, the entanglement
of a block of size $L$ with the rest of the chain  scales as \cite{lato04}
\beq
S(L)\simeq \frac{c}{3}\log L \ ,
\eeq
where $c$ is the central charge associated with the universality class
of the quantum phase transition. In particular, we can take 
party $A$ to be the left half of the chain with
$L=N/2$ sites and party $B$ to be the right half with the remaining sites.
It is clear that the entanglement of half of the chain with the other half
will diverge as $N$ goes to infinity. More precisely, if we consider
a system with open boundary conditions,  the following diverging behavior
is expected
\beq
S({\rm infinite\ half-chain})\buildrel{N\to\infty}\over{\longrightarrow} \frac{c}{6} \log \frac{N}{2} \ .
\eeq
Asymptotically, for very long chains, the half chain entropy is only half of the entropy of a  block  with the same size. This can be understood by noticing that the block has two boundaries available to establish correlations with the rest  of the chain, whereas a half chain only has one.
We may now wonder how much of this infinite amount of
entanglement is captured by the MPS approximation. 
For a system in an MPS with matrices of size $\chi$, $S(\rho_A)$ is 
trivially bounded by $\log \chi$. It is thus obvious that an 
MPS with matrices of finite size cannot describe exactly
the behavior of an infinite system \emph{at} the critical point but
we may try to find the exact amount of entanglement which
is captured. 

We have found that the quantitative 
entanglement support of MPS at
criticality obeys the following scaling law for the quantum Ising model
\beq 
S_\chi=\frac{1}{6}\log \chi \ 
\eeq
with a remarkably high precision.

This effective saturation of the entropy can be understood 
in an elegant way
as the emergence of a finite correlation length $\xi_\chi$,
a fact that was first analyzed in Ref. \cite{anderson99} in the
context of DMRG calculations for gapless systems.
To complete the connection we use 
the known result \cite{cardy-calabrese} that, near criticality, 
entanglement entropy is expected to be
saturated by $S \simeq \frac{c}{6}\log \xi$.  Typical values of the central charge are $c=1/2$ for the Ising model and $c=1$ for the Heisenberg model.

Thus, our result hints
at the {\sl finite-}$\chi$ {\sl scaling relation} 
\beq
\xi_\chi=\chi^\kappa\qquad {\rm with}\ \kappa=2,
\eeq
for the quantum Ising model. Moreover,
we shall find this relation to be fully consistent with 
many other scaling properties in the system. In some sense
we may argue that the finite matrix-size $\chi$ inherent
to the MPS approximation works as a probe of
the universality class of the quantum phase transition
which is investigated, a fact which is analogous to
the well-known finite-size scaling for finite systems \cite{Domb}.

Our results will be mainly numerical obtained with 
a specific technique. The best
MPS approximation to a given state can be obtained using
different algorithms, DMRG being the most popular choice.
Nevertheless, the recently proposed infinite time-evolving block decimation
\texttt{iTEBD} \cite{Vidal:infinite-chain,orusvidal07} turns out to
be particularly suited to address infinite
quantum systems. This algorithm exploits translational invariance,
makes the programming quite simple and, for our purposes, runs faster than 
the commonly used finite size DMRG. 
Yet, we have verified that the results we are presenting here
can be obtained using DMRG. We are therefore led to believe that our 
findings are intrinsic to the MPS representation and are not really 
sensitive on the precise algorithm used to get an approximation of the ground state.

We also have compared our results with the corresponding classical ones when available.
The agreement emerging from this comparison is a hint that the scaling properties we are 
facing of MPS could be a general phenomenon for quantum phase transitions studied with tensor network techniques
(of which the MPS is just a possible choice).

We would like to stress that our goal is to settle the scaling properties 
inherent to the MPS
approximation. For that purpose, we do \emph{not} need to work with MPS's 
with matrices of very large size $\chi$ as far as we reach the scaling 
region. This  region, for the case we study  is defined by 
\beq
\xi_{\chi} \gg a
\eeq
where $a=1$ is the lattice spacing. Hence, depending on the value 
of $\kappa$, the scaling region  can be attained with very modest values of $\chi$.

The paper is organized  as follows. In section \ref{sect:finitechiscaling} we discuss the 
origin of a finite-$\chi$ scaling relation. Then, in section \ref{sect:evidence}, we collect numerical evidence supporting its validity. In section \ref{sect:heisenberg}, we show 
that a similar scaling relation is expected for the Heisenberg model. Some applications of finite-$\chi$ scaling are briefly discussed in section \ref{sect:applications}. Namely, we will show how to extract critical exponents from finite-$\chi$ scaling. We summarize our results in section \ref{sect:conc}. Details regarding the  \texttt{iTEBD} algorithm, its convergence and some improvements we have implemented are presented in the Appendices.

\section{Finite $\chi$ Scaling}\label{sect:finitechiscaling}
%%%%%%%%%%%%%%%%%%%%%%%%%%%%

Phase transitions are usually detected through a local order parameter that discriminates between the two phases separated by the critical point. Let us consider a concrete example, the 
infinite quantum Ising model in a transverse field \cite{Sachdev}
\beq\label{eq:ising}
H=-\frac{1}{2} \sum_{i} \left (\sigma^{x}_i \sigma^{x}_{i+1} + \lambda \sigma^z_i \right).
\eeq
The phase transition of this model is driven by the transverse magnetic field, $\lambda$.
The $x$-magnetization plays the role of an order parameter and scales as $M \equiv \mean{\sigma^x_i} \sim |\lambda^2-\lambda^{*2}|^{1/8}$ near the critical point $\lambda^*=1$ \cite{pfeuty}.

We expect that, at criticality, a description of the ground state of $H$ in terms of a finite $\chi$ MPS blurs a phase transition smooth. For instance a diverging correlation length at $\lambda^*=1$ is replaced by a peak for the value of $\xi_{\chi}$ at some value $\lambda^*_{\chi}$ of the transverse field ($\lambda^*=\lambda^*_{\chi\to\infty}$). Indeed the correlation length of an MPS is usually finite \cite{ostlund,wolf05}\footnote{unless we are in the in the degenerate case of an mps phase transition due to nearing the two first eigenvalues of the transfer matrix  \cite{wolf05}. We have not observed this effect in  our simulations.}.

The value of the peak, $\xi_{\chi}$ and its position $\lambda^*_{\chi}$ should be dictated by a scaling relation of the following type
\beq\label{eq:xichi}
 \xi_{\chi} \sim \chi^{ \kappa} .
\eeq
Let us briefly argue why this should be the case, by showing how the arguments in Ref.\cite{Domb}, formulated for finite size scaling, can be adapted to the case of finite $\chi$ scaling. 
If Eq. (\ref{eq:xichi}) holds, in analogy with what is observed in finite systems,  the MPS finite  $\chi$  smooths all the divergences that  we would observe in infinite systems at the phase transition. They should be transformed to some finite anomaly at a $\chi$ dependent pseudo-critical point $\lambda^*_{\chi}$.
To see this, we start by  noticing that, asymptotically, the correlation length depends only on the distance from the transition through the  universal critical exponent $\nu$:
\beq\label{eq:xiti}
\xi \sim t ^{-\nu} ,
\eeq  
where $t=\vert \lambda-\lambda^*\vert / \lambda^*$.
By reading this relation  in the opposite direction we gain some further understanding
\beq \label{eq:txi}
t \sim \xi ^{-1/\nu} .
\eeq  
Given that $\chi$ cannot be taken to infinity,  we are keeping the system
away from criticality. The  transition  is actually shifted  to  a pseudo phase transition located at a different value
of the magnetic field
$\lambda_\chi^*$. There, the correlation length does not diverge. 
By substituting Eq. (\ref{eq:xichi}) into Eq. (\ref{eq:txi}) we obtain a prediction on how the 
pseudo-critical point should approach the true critical point when varying $\chi$:
\beq\label{eq:lambdashift}
\frac{|\lambda^*_\chi-\lambda^*|}{\lambda^*} \sim \chi^{- \kappa/\nu}.
\eeq
 
For a given $\chi$, we  obtain the effective distance from criticality when the system is at its  critical point. 
We can hence stick there , at $\lambda^*$, and  fix our attention on how  universal quantities  should vary as we  change $\chi$. 
We may now envisage three different scenarios.
When a universal quantity $F_u$ diverges  approaching the critical point with an exponent $\omega$ this translates to a divergence at $\lambda^*$ in term of $\chi$ as:
\beq\label{eq:univ-div}
F_u(\lambda^*) \sim \chi^{\frac{\omega  \kappa }{\nu}} .
\eeq
In the case where the universal quantity vanishes when approaching the critical point  with a given  exponent $\upsilon$, 
as is the case for the order parameter, then we should have
\beq\label{eq:univ-van}
F_u(\lambda^*)\sim \chi^{-\frac{\upsilon  \kappa }{\nu}} .
\eeq
As a last case, we consider the possibility of a  logarithmic divergence, as is the case for the half chain entropy. Then,
\beq \label{eq:univ-log}
F_u(\lambda^*)\sim  \kappa  \log (\chi) .
\eeq

Now we can look for deviations from the critical point. Once we have isolated the anomalous contributions to the universal quantities we are left with a regular part that,  if correctly interpreted, does not depend on the size of the matrices.  In analogy to the finite size case, we call this contribution  the scaling function for that particular  universal quantity.
An intuitive picture of its origin can be obtained by considering again Eq. (\ref{eq:xiti}).
We consider the variable 
\beq\label{eq:scal-var}
x=t\:\xi^{1/ \nu}
\eeq
that, in  an infinite system, stays of order one  in all the critical region, including the phase transition point  as guaranteed by 
Eq. (\ref{eq:xiti}). Away from the critical region, where the correlation length attains a finite value, it increases monotonically  with $t$.  When passing to finite $\chi$ systems  we break the relation Eq. (\ref{eq:xiti}). Expressing the correlation function in term of $\chi$ by using 
Eq. (\ref{eq:xichi})  we get
\beq\label{eq:scal-var-f}
 x=t \chi^{\kappa / \nu}.
\eeq
 
Values for this variable close to zero, are due to finite $\chi$ effects  and can easily 
be obtained by getting closer and closer to the critical point at fixed $\chi$.
This is the  variable that really quantifies the distance from an infinite system. 
Systems with different $\chi$ at different $t$ but with the same $x$ are  indeed at the same distance from the  corresponding infinite system. 
The variable $x$ can thus be used to unmask $\chi$ independent effects induced by forcing the system away from its critical behavior. 
In order to do this, however, one should keep in mind that systems with different $\chi$ have also different anomaly strengths as described by equations (\ref{eq:univ-div}), (\ref{eq:univ-van}) and (\ref{eq:univ-log}). In order to unmask $\chi$ independent effects  we should therefore  normalize the results obtained with system with different $\chi$  with their anomalous contributions at the transition.  For the cases considered in Eq. (\ref{eq:univ-div}), 
(\ref{eq:univ-van}) and (\ref{eq:univ-log}) the scaling functions are extracted respectively as
\beq\label{eq:sc-univ-div}
f_u(x) \sim \chi^{-\frac{\omega  \kappa }{\nu}} F_u (x) ,
\eeq
\beq\label{eq:sc-univ-van}
f_u(x)  \sim  \chi^{\frac{\upsilon  \kappa }{\nu}} F_u(x) , 
\eeq
\beq \label{eq:sc-univ-log}
f_u(x) \sim \frac{F_u( x )} { \kappa  \log (\chi)} .
\eeq

We  now provide numerical support to the finite-$\chi$ scaling.

\section{Evidence for finite-$\chi$ scaling for the quantum Ising chain}\label{sect:evidence}

The general discussion on finite-$\chi$ scaling should be verified on
concrete examples. We present in this section the results for the quantum Ising chain in a 
transverse magnetic field in Eq. (\ref{eq:ising}). All our results are obtained using the 
\texttt{iTEBD} algorithm. Some aspects of this technique are
discussed in the Appendix. 

\subsection{Half-chain entropy}
%%%%%%%%%%%%%%%%

We first  compute the von Neumann entropy for half the infinite
chain. As mentioned previously, this measure of entanglement should diverge with the
size of the system. Such a divergence cannot be accommodated by a finite-$\chi$ MPS ansatz.
Entanglement must circulate via the ancillary indices of the matrices that build the
approximation. For matrices of size $\chi$, entanglement is bounded to only span
a space of dimension $2^\chi$ as explained in Ref. \cite{vers05}, rather than the actual 
diverging $2^\frac{N}{2}$ dimensions. Moreover,
the eigenvalues in the Schmidt decompositions obey some decay law (an exponential decay,
up to degeneracies, is expected from conformal field theory), that further decreases
the amount of entanglement that the approximation should support.

\begin{figure}[hbt]
\includegraphics[scale=0.34]{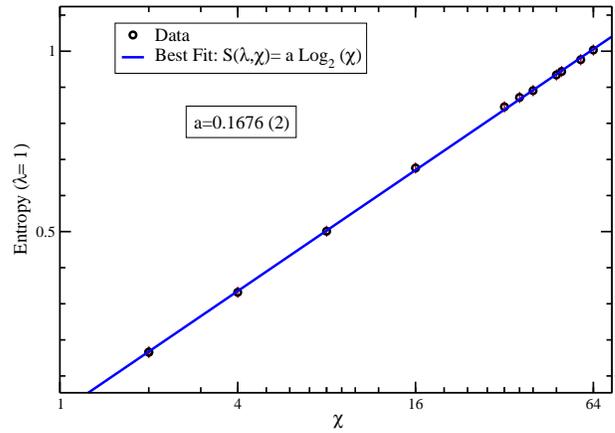}
\vspace{0.1cm}
\caption{Entropy as a function of $\log\chi$ at 
$\lambda=1$.}\label{fig:entropy-logchi}
\end{figure}

Numerical results for the entanglement entropy for the half-chain at $\lambda=1$ is
shown in Fig. \ref{fig:entropy-logchi}, 
where we have plotted $S_{\chi}$ as a function of $\chi$ and found an accurate fit to the scaling law 
\beq
 S_{\chi}\simeq\frac{1}{6} \log \chi .
\eeq
The remarkable precision of the fit should emerge from the absence of constant
and $\frac{1}{\log \chi}$ corrections. This effect was observed in the context of
block entropies in Ref. \cite{singlecopy} and shown absent in other 
measures of quantum correlations like the single copy entanglement which
is based on the largest eigenvalue of the reduced density matrix of a subsystem.
Conformal symmetry orchestrates a cancellation of subleading terms
coming from all the eigenvalues of that reduced density matrix. 
In the present case, it is unclear why corrections are absent 
in the computation of the half-chain entropy at
the point $\lambda=1$. 

\vskip 1cm

We can now match this scaling to the result of Ref.\cite{cardy-calabrese} that
states that, away from criticality, $S \simeq \frac{c}{6}\log\xi$. Then, the hypothesis 
of finite-$\chi$ scaling suggests the half-chain entropy to behave as
\beq
S_{\chi} \sim \frac{c}{6} \log \chi^{\kappa}.
\eeq
In the case of the Ising model, we find 
\beq\label{eq:val-k-ising-s}
\xi=\chi^\kappa,\qquad {\rm with\ } \kappa\simeq 2.011(2),
\eeq
where we have used that the central charge $c$ is equal to $1/2$
for the Ising model. The error in our result only reflects
the quality of the fit. This depends on the use of small values
of $\chi$, where scaling may not be present, 
and on possible violations of that scaling.
The uncertainty is then not representing 
a faithful systematic error but just the order of magnitude
of the freedom in the fit.
Our goal in this paper remains to collect a first consistent estimate 
for what is the actual value of $\kappa$.

In practical terms, this result shows that numerical exploration of the
critical properties should be well described using relatively small MPS.
A value of $\chi\sim 20$ describes faithfully correlations up to
400 sites.

We now consolidate this result by checking its consistency
with the computation of other observables.

\subsection{Shift of the critical point} 
%%%%%%%%%%%%%%%%%%

In the vicinity of the critical value $\lambda^*=1$, the entanglement entropy of half of the Ising chain diverges and the magnetization abruptly drops to zero. The best MPS approximation to
this scenario manages to produce a peak in the entropy and sudden drop of the magnetization
for values of $\lambda$ which are shifted from the infinite chain critical value. 
We label  $\lambda^*_{\chi,S}$ the coupling where the entropy presents a peak and the  $\lambda^*_{\chi,M}$ the one  where the magnetization vanishes abruptly. As in finite size simulation schemes \cite{Domb}, we have found that both  $\lambda^*_{\chi,S} \neq \lambda^*$ and  $\lambda^*_{\chi,M} \neq \lambda^*$. But we have found that, within the accuracy of our simulation, $\lambda^*_{\chi,M}=\lambda^*_{\chi,S}=\lambda^*_{\chi}$. This can
be understood as a check of the consistent representation of criticality that MPS develop.

\begin{figure}[hbt]
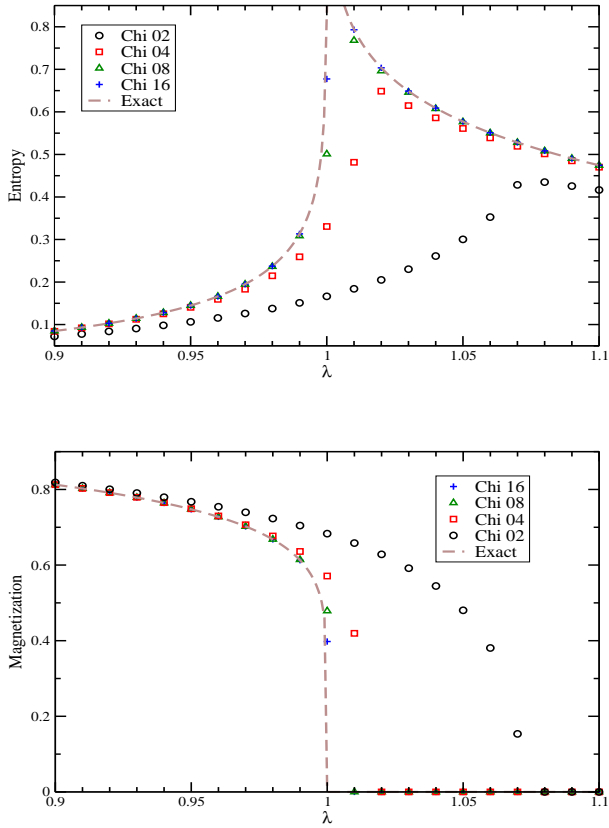

\includegraphics[width=8cm,height=5cm]{Graf-Entr-hz-090-120.eps}\null\vskip .9cm
\includegraphics[width=8cm,height=5cm]{Graf-Mx-hz-090-120.eps}
\caption{Entropy and magnetization around the theoretical critical point obtained for $\chi=2,4,8$ and $16$ using, as described in the appendices, $\varepsilon=10^{-1}$ and after convergence of the eighth decimal. The error bars due to the finite value of $\varepsilon$ are smaller than the points size.}\label{fig:Entropy}
\end{figure}

Our results are shown in Fig. \ref{fig:Entropy} for the entropy and the magnetization respectively. We can see that (i) the amplitude of the shift $\lambda^*-\lambda^*_{\chi}$ reduces when we increase $\chi$, (ii) the peak of the entropy rises with increasing $\chi$, and that (iii) far from the critical point, modest values of $\chi$ are sufficient to get faithful approximation of the ground state (in the sense that the curves obtained for different values of $\chi$ tend to collapse). 

\begin{figure}[!hbt]
\includegraphics[scale=0.34]{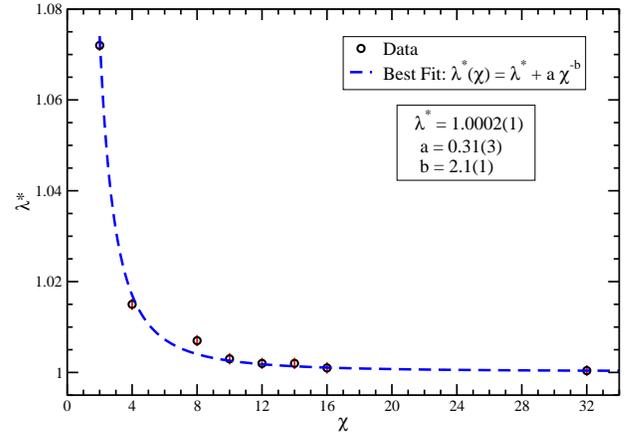}
\vspace{0.1cm}
\caption{Effective critical point $\lambda^{*}_{\chi}$ as a
function of $\chi$. As discussed in the appendices, the values for $\chi=2$ and $4$ were obtained with 
$\varepsilon=10^{-3}$, while for $\chi=8$, $16$ and $32$ with $\varepsilon=10^{-2}$. The errors bar are due to the finite value of $\varepsilon$.}\label{fig:Critical Point}
\end{figure}

We have checked that the shift of the critical point obeys the law (\ref{eq:lambdashift}). The results  are plotted in Fig. \ref{fig:Critical Point}. As expected, the way  $\lambda^*_{\chi}$ approaches $\lambda^*$  is correctly described by  a power law. Using $\nu=1$ we extract:
\beq\label{eq:val-k-ising-lc}
\kappa=2.1(1)
\eeq
where, again, the error is only reflecting the precision of the fit.  

This value is compatible with the value extracted using the entanglement entropy. 
We see, however, that this estimation is less precise. This fact is related to the difficulties encountered in the determination of $\lambda^*_{\chi}$. In principle the finer the scan,  the more precise the value of  $\lambda^*_{\chi}$.  However the sharpness of the scan is limited by the numerical precision with which we obtain the entropy. At some point, entropies of chains with close but different values of $\lambda$ are compatible within their error bars. Then, we cannot  further refine our scan and should accept the obtained precision as the best we can achieve for the location of the transition.\vskip .5cm\null

\vskip 1cm
\vskip .2cm

\vskip .2cm

\subsection{Magnetization}
%%%%%%%%%%%%%% 

\begin{figure}[hbt]
\includegraphics[scale=0.34]{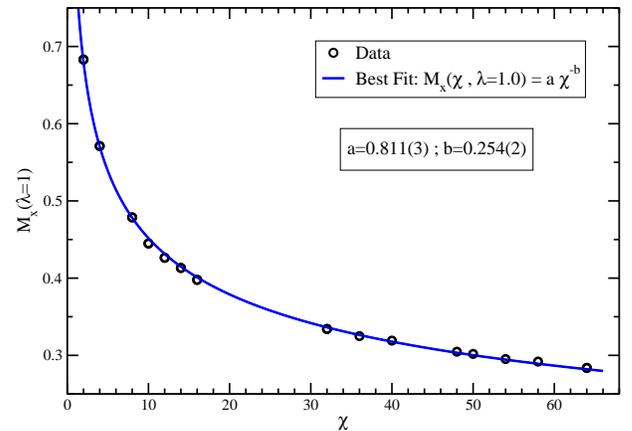}
\vspace{0.1cm}
\caption{Magnetization as a function of $\chi$ at  $\lambda=1$. }
\label{fig:Mx-Chi} 
\end{figure}

The drop of the magnetization near the critical point obeys scaling laws as discussed
previously. 
 We actually expect the magnetization at finite $\chi$ to behave as
 $M_{\chi}(\lambda=\lambda^*)\sim \chi^{-\frac{\beta  \kappa }{\nu}}$ with the Ising critical exponents
  $\beta=1/8$ and $\nu=1$. We may now take 
our numerical results and fit $\kappa$ in this expected scaling law.
 In Fig. \ref{fig:Mx-Chi}, we have plotted $M_{\chi}$ as a function of $\chi$ for the Ising chain at 
$\lambda=1$. By fitting our numerical results with a function of the form $a \chi^ b$ (see Fig. \ref{fig:Mx-Chi}), we obtain:
\beq\label{eq:val-k-ising-m}
\kappa=2.03(2).
\eeq
This value of $\kappa$ is in agreement with  our two previous determinations. 

\vskip 1cm

\subsection{Block entropy} 

A new consistency check consists in  considering the entropy of the reduced density operator of a block of $L$ contiguous particles. For a critical systems, this entropy scales with $L$ as $S_L \simeq \frac{c}{3} \log L$ \cite{lato04}. We have observed that, for a fixed value of $\chi$, this entropy saturates at a distance $L \simeq \chi^ {\kappa}$. We can make a very qualitative assumption on the fact that the length at which the entropy saturate is of the order of the correlation length and in this way use this value as determination of the correlation length. It is likely that this qualitative assumption can be made rigorous in a renormalization group framework  by introducing a new scaling field $\chi^{\kappa / \nu}$. However this analysis is out of  the scope of this paper.
By using the relation (\ref{eq:xichi}) with this estimation of the correlation length we can have an idea of the magnitude of $\kappa$.

In order to compute the entropy of a block of $L$ spins, we have used the ideas contained in the work by Verstraete et al. \cite{vers05}. The basic idea is to reconstruct the effective new matrix MPS upon 
successive RG coarse-graining transformations.
Our results are displayed on Table \ref{tab:Block Entropy}, where we can see that $S_L$ saturates for $L \simeq \chi^2$. So that we get a further confirmation that, for the Ising model,
\beq
\kappa\sim 2.0(1) ,
\eeq
in agreement with the previous estimations, though less accurate. We observe that  for sufficiently large $L$, $S_L$ is approximately equal to two times $S_\chi$  (the half-chain von Neumann entropy) calculated at the same $\lambda$. Let us recall that the explanation for this factor $2$ is that a finite block has two boundaries available to establish correlations with the rest  of the chain, whereas a half-infinite chain only has one.

\begin{table}[ht]
\begin{tabular}{c c c c}
$L$  & $S(L,\chi=2)$ & $S(L,\chi=4)$ & $S(L,\chi=8)$
\tabularnewline \hline \hline 
2    &  0.2994  &  0.4825  &  0.5883  
\tabularnewline 
4    &  0.3279  &  0.5647  &  0.6976
\tabularnewline 
8    &  0.3317  &  0.6271  &  0.7934
\tabularnewline 
16   &  0.3317  &  0.6586  &  0.8720
\tabularnewline 
32   &  0.3317  &  0.6586  &  0.9288
\tabularnewline 
64   &  0.3317  &  0.6586  &  0.9577
\tabularnewline 
128  &  0.3317  &  0.6586  &  0.9630
\tabularnewline 
256  &  0.3317  &  0.6586  &  0.9632
\tabularnewline 
512  &  0.3317  &  0.6586  &  0.9632
\tabularnewline 
1024 &  0.3317  &  0.6586  &  0.9632
\tabularnewline \hline \hline 
\end{tabular}
\caption{Entropy of a block of $L$ spins  using the ideas contained in \cite{vers05}. We observe that the entropy saturates around $L\sim \chi^2$. Note that the values obtained for the entropy after saturation are the double of those  obtained for half of the chain. This factor of two is due to the fact that here the block has two boundaries.}\label{tab:Block Entropy}
\end{table}

\subsection{Correlation length}\label{sec:scp}
%%%%%%%%%%%%%%%%%%%%%%%%%

\begin{figure}[hbt]
\includegraphics[scale=0.34]{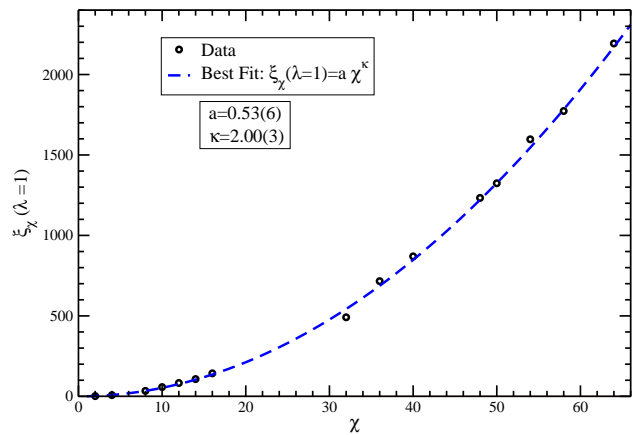}
\caption{Correlation length as a function of the size of the $\chi$ in the case of the Ising model at $\lambda=1$.}\label{fig:xichi-Ising}
\end{figure}

All our previous results should be a consequence of the emergence of a finite
correlation length $\xi_\chi$. This fact was first investigated in
Ref. \cite{anderson99}. We can address this point by analyzing
the ratio of the two highest eigenvalues of the transfer matrix \cite{wolf05}
computed from the matrices in the MPS. On Fig.(\ref{fig:xichi-Ising}), we have plotted the value of $\xi_{\chi}$ as a function of $\chi$. To extract the value of the exponent $\kappa$, we have performed a fit to numerical data with a function of the type $a \chi^ \kappa$ with $a$ and $\kappa$ left as free parameters. We have found 
\beq \label{eq:val-k-ising-xi}
\kappa=2.00(3).
\eeq
Again, the consistency of this result with our previous determinations
is manifest.

\vskip .6cm

\subsection{Scaling function for the magnetization}\label{sec:sfm}

A further manner to test finite-$\chi$ scaling is to analyze in more detail the magnetization.
It follows from the scaling analysis in Sect. \ref{sect:finitechiscaling} that 
$M_\chi$ depends on $\chi$ only through the product $x=\chi^{ \kappa/\nu} t$. 
Therefore,  we can plot the rescaled magnetization 
$M_{\chi}(\chi^{ \kappa/\nu} t) \chi^{ \kappa \beta/\nu}$ 
as a function of $\chi^{ \kappa/\nu} t$ for different values of $\chi$,
assuming the known values of $\nu=1$ and $\beta=1/8$ of the Ising universality class. In case finite-$\chi$
scaling is verified, all points should lie on the same curve. The quality of this collapse is, hence,
a function of the correct value of $\kappa$ alone. 

We have  scanned $\kappa$ for a broad range of values and selected the ones  that qualitatively produced  a collapse of the numerical points onto a single curve. 
Remarkably, we have verified that only for a relatively small interval of $\kappa$ values,  
all the  points obtained with this procedure  lie on the same curve. 
Whatever small variation outside this interval of $\kappa$ reflects on a sensible spread of the point outside the curve.

\begin{figure}[hbt]
\includegraphics[scale=0.34]{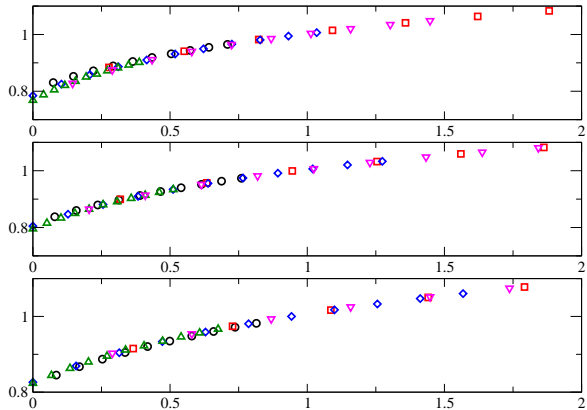}
\vspace{0.30cm}
\caption{Collapse of the rescaled magnetization obtained with different MPS using distinct values of $\kappa$ on the scaling function: $\kappa=1.9, 2.0, 2.1$ for  upper, middle and lower graphs, respectively. }\label{fig:FSS}
\end{figure}

Our results are displayed on Fig.\ref{fig:FSS}. Again, we find a further confirmation that
$\kappa\simeq 2.0(1)$ is the right scaling exponent.

\subsection{Scaling function for the the energy difference and comparison with classical results}\label{sec:sfe}

 In this section, we clarify why studying the deviation from the critical point with finite MPS we face a two scale problem. This problem is the quantum version of the already studied  two scale  finite-size scaling ansatz in the context of classical systems  \cite{nishino}.
In our case, the analysis is performed  considering infinite size systems.The first scale is given by the MPS dimension   $\chi$  and the second scale  is given by $t$. We saw in the previous section that by correctly treating the two scales, one is able to extract universal scaling functions. The universality, implies that  scaling functions however, should not  depend on which system, among those in  the same universality class, one decide to consider.
This statemet can be checked  by comparing our results with the one contained
in Ref. \cite{nishino} about the classical Ising model in two dimensions at critical temperature. This system is indeed in the same universality class we are considering: the two dimensional Ising universality class. 
The authors  apply the ideas of the Corner Transfer Matrix (CTM)  renormalization group  \cite{CTMRG} to it.
This is a technique that generalizes the DMRG  renormalization ideas and its related variational techniques over  MPS to a real space renormalization algorithm for classical systems. 
Once the CTMRG is applied at critical temperature as in  ref. \cite{nishino} a new scale   emerges. This is  the inverse of a correlation length depending on the dimension of the renormalized CTM $m$. The authors label  this scale  $\xi(m)$. This scale exactly corresponds to the scale we are calling here $\xi_{\chi}$ . In this  way,  the authors of ref. \cite{nishino},  by treating the finite size classical system of dimension $N$ studied with a finite renormalized CTM  of dimension $m$ as a two scale problem,  extract the value of all critical exponents. 
 Here we consider the precise map between our results and the one contained there.
Following  the recipes in Ref. \cite{CTMRG} we see  that, (as already implicit in the identification of scales) the classical correspondence of $\chi$ is the size of the renormalized CTM (called $m$ in both references).  We can use again  the finite size scaling ansatz and map the distance from the critical point that we call $t$ to the size of the classical system  considered in ref. \cite{nishino} ( called $N$ ). In this way the scaling variable $x_m=\xi(m)/N$ of ref. \cite{nishino} is related to the one we use $x$  as : $x_m=x^{\nu}$. As for the Ising model $\nu=1$, the results in ref. \cite{nishino}  should exactly correspond to ours.
We check this claim by consider what in our language would be the plot in figure 3 of Ref. \cite{nishino}. It represent the energy difference as a function of $x$ with respect to the exact result. We can use the standard mapping between the free energy of a classical system and the ground state energy of the corresponding quantum system and compare the plot in figure 3 of ref. \cite{nishino}. with our results for the ground state energy per bond.
To do this we  plot the difference of the ground state energy with a given $\chi$ and the exact result $E(\infty)=-1.27323954$. The scaling function for the energy difference is obtained  by plotting  $(E_{\chi}(1/x)-E_{\infty}) \chi^{ \kappa/\nu}$  \footnote{We warn the reader of a misprint in fig. 3 of Ref. \cite{nishino} where both the caption and the x axis label should read $N/\xi(m)$ as explained in the reference main text instead of $\xi(m)/N$ as written.}. Again we see that by using a value of $\kappa=2$  the points obtained with different $\chi$ collapse to a single curve.
We also see that the curve we obtain has the same shape but a different normalization factor  with respect to the one obtained in figure 3 of  Ref. \cite{nishino}. This is expected since the normalization factors are known to be universal but boundary condition dependent \cite{Fisher} and we used different boundary condition  from  the ones used in Ref. \cite{nishino}.

\begin{figure}[hbt]
\includegraphics[scale=0.34]{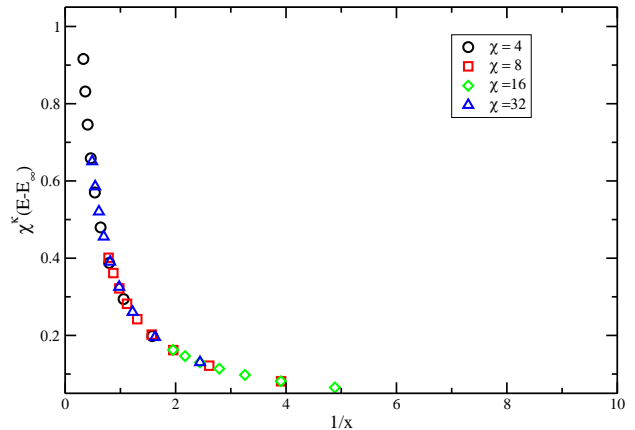}
\vspace{0.30cm}
\caption{Collapse of the rescaled energy difference  obtained with different MPS. The values we use are  $\kappa=2$, $E(\infty)= -1.27323954$.}\label{fig:FSSE}
\end{figure}
We can hence confirm that the scaling observed in this work in the case of a quantum phase transition is the analogue to the one observed in a classical phase transition in ref. \cite{nishino}.

\section{Evidence of finite-$\chi$ scaling for the Heisenberg chain}\label{sect:heisenberg}

An extensive analysis of the emergence of finite-$\chi$ scaling 
in different models is necessary to gain insight in the role
of the scaling exponent $\kappa$. Here, we only make a 
first step and explore the 
Heisenberg spin 1/2 Hamiltonian
\beq
H=\sum_i \vec\sigma_i\cdot\vec\sigma_{i+1}  \ .
\eeq
We may conjecture that $\kappa$ should   only vary with 
the universality class of the model considered.
To assess the new value of $\kappa$, we consider the scaling of the half chain entropy since
this strategy provided very precise determination in the Ising case.

\begin{figure}[hbt]
\includegraphics[scale=0.34]{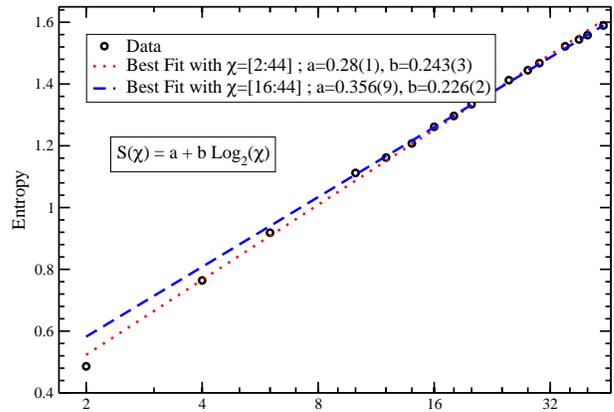}
\vspace{0.1cm}
\caption{Entropy as a function of $\log\chi$ for the Heisenberg model.  Data have been fitted with a function of the type $a+b\log (\chi)$ with $a$ and $b$ free parameters. The results of the factor $b$ for the  fit in the  $\chi$ interval from $16$ to $44$ is $b=0.226(2)$.\\ \\} 
\label{fig:entropy-logchi-heis}
\end{figure}

We then follow the same steps as described for the Ising case and
we take the central charge to be $c=1$.  By fitting the numerical 
data with a curve of the type $a+b \log \chi$   and using 
the actual value of the central charge we obtain, as observed on Fig. \ref{fig:entropy-logchi-heis}, 
\beq\label{eq:val-k-heis-s}
\kappa=1.36(2).
\eeq
Let us note that the fit now includes a non-zero intersect. This was
absent in the Ising case. 

\begin{figure}[hbt]
\includegraphics[width=8cm,height=5cm]{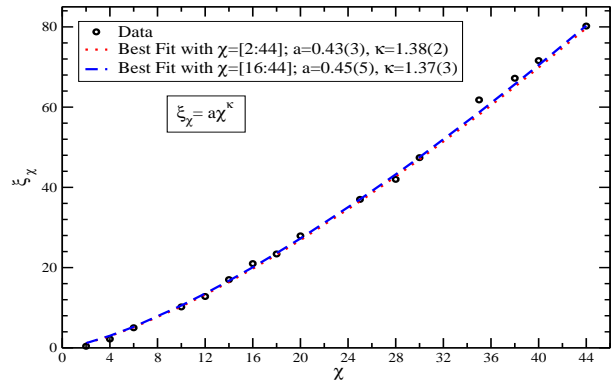}
\caption{Correlation length as a function of  $\chi$ in the case of the Heisenberg model.
This behavior can be correctly described by a relation of the type \ref{eq:xichi} with an exponent $\kappa=1.38(4) $. The fit has been performed in the $\chi$ interval $20 - 44$}
\label{fig:xichi-heis} 
\end{figure}

This result can be checked for consistency in a similar way as the results presented
for the Ising model. Here, we present as a further piece of evidence
for finte-$\chi$ scaling the scaling of the correlation length
as computed from the ratio of the largest eigenvalues of the transfer matrix.
As shown in Fig. \ref{fig:xichi-heis}, the numerical data  are described correctly by a
law of the type in Eq. (\ref{eq:xichi}) with an exponent
\beq \label{eq:val-k-heis-xi}
\kappa=1.38(2).
\eeq
\vskip 1cm

\vskip 1cm
\null
\vskip 1cm

Both determinations in Eq. (\ref{eq:val-k-heis-s}) and (\ref{eq:val-k-heis-xi}) 
are compatible   and support the value $\kappa\sim 1.37(2)$, which depends on the universality class of the model under discussion\footnote{A result compatible with our determination of $\kappa$,  has been obtained independently by R. Davies and R. Or\'us in a simulation of the superfluid phase of the Bose Hubbard model. They  confirmed  this value to us in a private communication. This constitute  a further hint that $\kappa$ is universal as all the superfluid phase of the Bose Hubbard model is critical and has $c=1$. }. 

\section{Applications of finite-$\chi$ scaling}\label{sect:applications}

As in the case of finite size scaling, we can use finite
$\chi$-scaling to extract critical exponents. The ideal strategy
is the one that does not rely on the knowledge of the position of the finite $\chi$
pseudo critical point. This is so because  the determination 
of the pseudo critical point is very delicate. Any small error in it propagates
to the determination of critical exponents as we explicitly saw when dealing with the determination of $\kappa$.

Keeping this in mind, we can envisage two different
scenarios: a first simple scenario, as the one of the Ising
model, when we know a priori the location of the phase transition.
In this case, in order to extract the critical exponents we
proceed as follows: i)Extract the value of $\kappa$ by  studying the behavior of $\xi_\chi$ at the
 critical point. ii) Extract all the ratios $\alpha
/\nu$ where here $\alpha$ represents a generic critical exponent by studying universal quantities as function of
$\chi$ at the  phase transition using  the value of $\kappa$ obtained in i).
iii) Extract the value of $\nu$ (and hence $\alpha$ from the ratio obtained in ii) )by
studying the derivatives of the universal quantities with respect
to $t$.
The second scenario and by far the  most unfavorable and  frequent is the one where
we do not know the location of the  critical point. In this
case, we need to adapt some strategy known from finite size analysis to extract its value
if we want to apply finite $\chi$-scaling to the transition.
A possibility is obtained by considering   the techniques of \cite{Nightingale} (more efficient methods can be found  i.e. in Ref. \cite{Marco1,Marco2}). A review of this method for the case of finite size scaling is contained in ref.  \cite{Vicari}.
We adapt it in the following way: we iteratively
obtain  estimates of $\kappa$ and the critical point by
considering the behavior of the correlation length as a function of
increasingly big $\chi$. 
Once these estimates converge to a fixed value, we can use the obtained values for $\kappa$ and for the critical point to repeat the steps
from i) to iii) of case one. In this way  we extract all the other critical exponents.
 The main source of error in all these determinations is, as in the
case of the finite size scaling,
the existence of scaling violations that we do not analyze in
this work. However, even without taking the scaling violations into
account, we think that the extracted exponent should be much more
accurate than the ones obtained with standard techniques.
 To justify this statement,  we review what we mean by standard techniques for extracting critical exponents with an infinite
MPS by considering again  the case of the Ising model.

\begin{figure}[hbt]
\includegraphics[width=8cm,height=5cm]{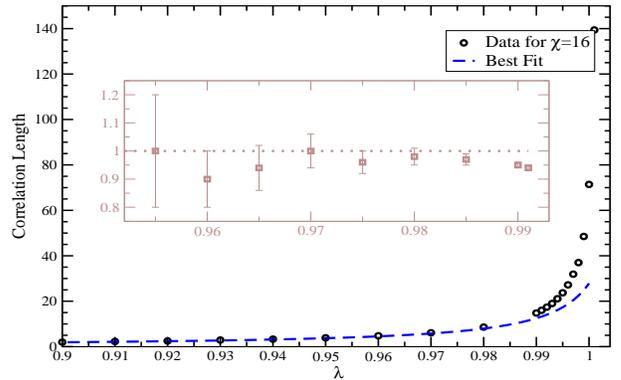}
 \caption{Fine tuning of the correlation length  around the
critical point for $\chi=16$ and $\varepsilon=0.1$. Note that the
points are not equally spaced. Inset: Values obtained for $\nu$ by
fitting magnetic field window of different sizes (all starting at
$\lambda=0.90$ and finishing at the point in the x-axis). 
We  notice a good region of stability which
can be used to extract our best estimate for the $\nu$
exponent.}\label{fig:fine tunning cl}
\end{figure}

We can extract the value of the exponent $\nu$ by
studying  the behavior of the correlation length at fixed $\chi$
when we approach the phase transition. We expect that far enough
from the region where finite-$\chi$ effects appear, a modest value
of $\chi$ should provide a faithful description of the Ising
ground  state.  The correlation length should obey a law
of the type $(\lambda -\lambda^*)^{-\nu}$. Fitting the data with
this function and  leaving $\lambda^*$ and $\nu$ as free
parameters we obtain an estimate of both $\lambda^*$ (the  phase 
transition point) and $\nu$. We also expect that due to systematic
errors induced by the fitting procedure (the difficult point is to
locate the correct window of $\lambda$ values for which we should perform
the fit) these estimates would have a slight dependence on $\chi$
and should converge to the exact $\lambda^*$ and $\nu$ for $\chi$ large
enough. In Fig. \ref{fig:fine tunning cl} we show the results
of such study, again for the Ising model. We extract as best estimate of $\nu$ in
the case of $\chi=16$
\beq\label{eq:nu} \nu\simeq 1.00(5) .\eeq
See also Ref. \cite{bursill,nishino,croo} and references therein to see how in the case of finite chains described with MPS, finite size scaling can be used as an alternative to extract critical exponents .

A similar strategy can be used to extract  the $\beta$ critical
exponent. Again, working slightly away from criticality, the
scaling of the magnetization is very nicely fitted with
\beq\label{eq:beta} \beta\simeq .1250(1) \eeq
 as shown in Fig. \ref{fig:fine tunning mx}.

\vskip 1cm
\begin{figure}[hbt]
\includegraphics[width=8cm,height=5cm]{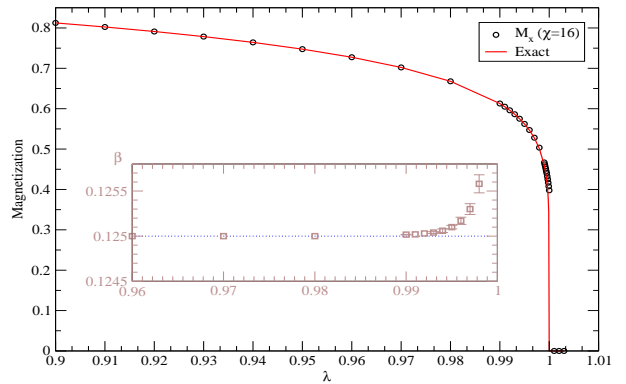}
\caption{Fine tuning of
the magnetization around the critical point for $\chi=16$
and $\varepsilon=0.1$. Note that the points are not equally spaced.
Inset: Values obtained
for $\beta$ by fitting magnetic field window of different sizes all
starting at  $\lambda=0.90$.
We clearly notice a good region of stability which can be used to
extract our best estimate for the $\beta$ exponent.}\label{fig:fine
tunning mx}
\end{figure}

In addition to the exponent $\beta$, one can consider the
exponent $\eta$ by studying the behavior of the two point
correlation function of the order parameter $\sigma_x$. Both
exponent are related via the hyper scaling relation: $ \beta=
(d-2 +\eta)/2$, where in this case $d=2$ as we are considering
the universality class of the classical two dimensional Ising
model.

\begin{figure}[hbt]
\includegraphics[width=8cm,height=5cm]{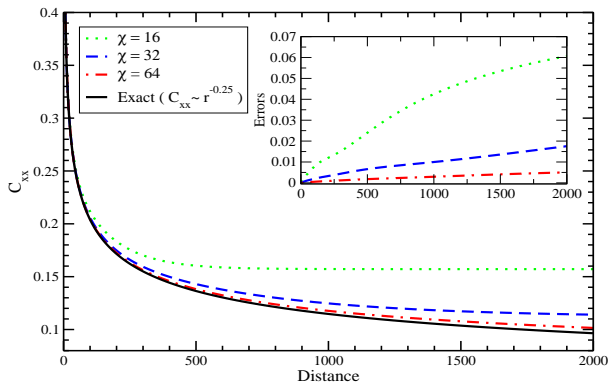} 
\caption{Study of the order parameter two point correlation function 
at $\lambda=1$ for $\chi=16,32,64$ and $\varepsilon=0.01$ compared with the expected exact behavior
$r^{-0.25}$. We note that the range of distances for which there is good agreement between the numerical correlation function and the exact result increases with $\chi$ as expected. Inset: Results of  fits with a power law of the type $a r^{-\eta}$ for the case of $\chi=64$ in the $r$ windows for which the extracted correlation functions  agree with the analytical results.}\label{fig:corr func}
\end{figure}

This relation implies that if $\beta=1/8$, then $\eta$ should be $1/4$.
We checked this for consistency. We plot the two point correlation
function of the order parameter as a function of the distance in
Fig. \ref{fig:corr func}. In a log-log plot, an algebraic decay
such $r^{-\eta}$ is seen as a straight line. We plot this straight
lines together with the correlations functions obtained for the
MPS at the phase transition with $\chi=16,32,64$. We appreciate how
the range for which the correlations reproduce the exact result
increases with the matrix dimension. Once the range of distances 
is correctly selected, a fit to a power law in the case of correlation
function of the $\chi=64$  MPS at $\lambda=1$ produce the
following best estimate for $\eta$ 
\beq\label{eq:eta} 
\eta \simeq .24800(25). 
\eeq
Again, this result only reflects the quality of the fitting strategy.
\\
\vskip 1cm

\section{Conclusions}\label{sect:conc}
%%%%%%%%%%%%%%%%%%%%

The amount of entanglement supported by the MPS approximation is limited by the size $\chi$ of the matrices that form the
ansatz. We have studied numerically this issue and found that all observables we have considered approach 
their exact values at criticality obeying scaling laws in $\chi$. The case of the 
quantum Ising chain in a transverse field is consistently described by an effective
finite correlation length that scales as $\xi_\chi=\chi^\kappa$, with $\kappa\simeq2$.
Most of the results presented here were related to the Ising model, but the numerical work we have performed shows that our findings are qualitatively valid for other models, such as the Heisenberg model where our calculations indicate that $\kappa \simeq 1.36$. 
Interestingly, the value of $\kappa$ seems to be model-dependent.

In the case of the Ising model, it is specially interesting to note the accurate fit of the half-chain
entropy to $S\sim \frac{1}{6}\log\chi$ at $\lambda=1$ with no constant or important 
subleading corrections. This effect is not present for  the Heisenberg model.

All our numerical results were found using the	\texttt{iTEBD} algorithm and checked to agree 
with standard DMRG \cite{mcculloch,dukelsky}.  It would be, in principle, possible
to use other algorithms as a brute force minimization of energy in the space of matrices in the MPS
structure. Such an approach may fail due to the proliferation of
local minima. Somehow, DMRG and	\texttt{iTEBD} manage to circumnavigate local minima an find
the absolute minimum within the approximation.

We have also checked that the scaling we encounter here coincides with the emergence of a second scale in some treatment of classical phase transition as pointed out in Ref. \cite{nishino}.

This correspondence is a hint that this phenomenon is quite general and appears whenever one tries to approximate operators with an infinite rank (such as the CTM or the half chain reduced density matrix) with finite rank operators 
Therefore it is likely that scaling  is not strictly related to the MPS representation of the ground state. 
We are currently investigating this issue by repeating a similar  study  to the one presented here with different tensor network representations \cite{arbol}.

With the same reasoning, we expect finite-$\chi$ scaling to appear 
for some  generalizations of MPS, such as Tensor Product States \cite{nishino2} also known as Projected Entangled Pairs States
\cite{two-dim}.
It remains an open problem to derive the scaling relation analytically for exactly solvable models.

\section{Acknowledgments}

We thank P. Calabrese, J. I. Cirac, J. J. Garc\'ia-Ripoll, Ll. Masanes,  
S. Montangero, R. Or\'us,  M. Roncaglia, E. Vicari and G. Vidal for discussions and suggestions on the topics presented here. 
We thank I. P. McCulloch for his comments on the manuscript.
Financial support from QAP (EU), MEC (Spain), Generalitat de Catalunya and CAPES (Brazil) is acknowledged.

\appendix

\section{Error control and convergence issues with the \texttt{iTEBD} algorithm}\label{sect:iTEBD}
%%%%%%%%%%%%%%%%%%%%%%%%%%

In this section, we wish to address the reliability of the data output by the \texttt{iTEBD} algorithm. Let us start by reminding the main features of this algorithm. A more technical presentation can be found in \cite{Vidal:infinite-chain}.

The \texttt{iTEBD} algorithm aims at finding the ground state energy per particle of a Hamiltonian of the form
\beq
H=\sum_{i=-\infty}^{\infty} h_i,
\eeq
where $h_i$ represents a two-spin next-neighbor interaction term. This algorithm is based on the following identity, valid for any gapped Hamiltonian:
\beq\label{eq:imaginarytimeevolution}
\ket{\Psi_g}=\mathcal{N} \lim_{\tau \to \infty} e^{-\tau H} \ket{\Psi_0}.
\eeq
That is, a ground state of $H$ can be obtained by evolving some initial state $\Psi_0$ in imaginary (Euclidean) time whenever $H$ has a gap above the ground state and  $\braket{\Psi_0}{\Psi_g} \neq 0$. For many Hamiltonians of interest, though, Eq.(\ref{eq:imaginarytimeevolution}) cannot be used as such. Rather, one computes the following sequence until convergence is attained:
\beq\label{eq:basicsequence}
\Psi_{i+1}=\mathscr{E}_{i}(\epsilon,H) \Psi_i/|| \mathscr{E}_{i}(\epsilon,H) \Psi_i ||,
\eeq
where $\epsilon$ is some tunable parameter such that  $\mathscr{E}_{i}(\epsilon) \simeq e^{- \epsilon H}$ for $\epsilon$ small enough. In the \texttt{iTEBD} algorithm, $\mathscr{E}_{i}(\epsilon,H)$ is decomposed into
\beq\label{eq:threeerrors}
\mathscr{E}_{i}(\epsilon,H)=\mathscr{Q}_{i} \mathscr{P}_{i} \mathscr{F}_i(\epsilon,H),
\eeq
where the factors appearing in the last expression correspond each to a different approximation that makes numerical computations tractable:
\begin{itemize}
\item i)
The first factor $\mathscr{F}_i(\epsilon,H)$ comes from using a cut off Suzuki-Trotter expansion \cite{Vidal:first} in order to approximate the action of $e^{- \epsilon H}$ by a product of two-body operators. (As a result, the form of   $\mathscr{F}_i$ depends on $i$.) The error introduced by truncating the Suzuki-Trotter expansion vanishes when $\epsilon \to 0$. We call this error {\sl finite time step error}. 

\item ii)
The second factor $\mathscr{P}_{i}$ is a projector that approximates $\mathscr{F}_i(\epsilon,H) \Psi_i$ by an MPS with matrices of some prescribed finite size $\chi$. This approximation is made in order to have an efficient description of the state at each step of the sequence (\ref{eq:basicsequence}). Indeed,  both storing of $\Psi_{i+1}$ and the computation of the mean value of a local operator now takes a time that is polynomial in $\chi$ \cite{Vidal:infinite-chain}. This approximation boils down to limiting the amount of correlations present in the system. We will call {\sl truncation error} the error due to this approximation. 

\item iii)
The operator $e^{- \epsilon H}$ is not unitary, and as a result $\mathscr{P}_{i} \mathscr{F}_i(\epsilon,H)$ neither is. This non-unitarity have small spurious effects that we can safely neglect \footnote{ Non-unitary gates may result in a loss of orthonormality between Schmidt vectors for bipartitions away from the two spins acted upon by the gate. See \cite{orusvidal07} for a recent discussion}. The third operator, $\mathscr{Q}_i$, does exactly this job, producing what we
call an {\sl orthonormalization error}. 
\end{itemize}

In order to study the time-step error, we have applied the \texttt{iTEBD} algorithm to obtain an MPS approximation of the ground state of the quantum Ising chain with matrices of size $\chi$ equal to 2. The reason why we have chosen to discuss the time-step error with such a small value of $\chi$ is that it is most illustrative. For various values of $\epsilon$ ranging from $10^{-1}$ to $10^{-5}$, we have computed the behavior of the ground state energy and the half-chain von Neumann entropy. It is natural to test the performance of the algorithm looking at the ground state energy since it is designed to minimize this quantity. It is less obvious why we also looked at the half-chain entropy. We will explain it shortly.

In principle, the smaller $\epsilon$, the more accurate the description of the state. But small values of 
$\epsilon$ also increase the number of time steps necessary to guarantee convergence of the simulation. One way to proceed, in order to correctly choose $\epsilon$, is as follows: (i) run a simulation with a rather large value of $\epsilon$, $\epsilon_1$, and get an estimate of the energy and the entropy. (ii) Repeat the simulation with a smaller value of $\epsilon$, $\epsilon_2$, and compare the resulting energy and entropy with those of the simulation at $\epsilon_1$. (iii) If the results are close enough (according to a predetermined margin), stop the simulation. Otherwise, repeat with smaller values of $\epsilon$ until convergence is attained.

\begin{table}
\begin{tabular}{c c c c c}
$\lambda$ & S1-S5 & S2-S5 & S3-S5 & S4-S5 \\
\hline\hline
0.9  &  9124  &   921  &    91  &    8 \\
1.0  & 42495  &  4203  &   416  &   38 \\
1.071& 368647  & 35489  &  3501  &  318 \\
1.072& 69632  &  6951  &   689  &   62 \\
1.073& 69200  &  6908  &   684  &   62 \\
1.1  & 58718 &  5871  &   581  &   53 \\
\hline\hline
\end{tabular}
\caption{Convergence of the 
entropy as a function of $\varepsilon$ for some values of $\lambda$.
The table shows the difference of the entropy found using a given  $\varepsilon$
with our best simulations corresponding to $\varepsilon=10^{-5}$ (
S1 is the value of the half-chain entropy obtained for $\varepsilon=10^{-1}$, S2 the values
obtained for $\varepsilon=10^{-2}$ and so on). All entries in this table should be multiplied by 
$10^{-8}$.}\label{tab:convergeps}
\end{table}

\begin{table}
\begin{tabular}{ c c c}
$\lambda$ & $\Delta E$ & $\Delta S$ \\
\hline\hline
0.5 &  $< 0.1$ &    39 \\
0.7 &  $< 0.1$ &   706 \\
0.8 &  $< 0.1$ &  2521\\
1.0 &  7     & 42495 \\
1.071 & 41 &  368647 \\
1.072 & 34 &   69632 \\
1.073 & 34 &   69200 \\
1.1 & 29 &     58718 \\
1.4 &  7 &     14226 \\
1.5 &  4 &      9807 \\
\hline\hline
\end{tabular}
\caption{$\Delta E$ and $\Delta S$ corresponding to the difference between the values
obtained for the entropy and energy when using $\varepsilon=10^{-1}$ and
$\varepsilon=10^{-5}$ for $\chi=2$. This gives an estimation of the error
due to $\varepsilon$ when using $10^{-1}$ as its value. Note
that the errors increase around the critical point and that the errors
in the entropy are much greater than the ones in the energy. All
entries of this table should be multiplied by $10^{-8}$.}\label{tab:erroreps}
\end{table}

On Table \ref{tab:convergeps}, we report on the convergence of the von Neumann entropy, as a function of $\epsilon$, for various values of $\lambda$, while Table \ref{tab:erroreps} shows the difference of the values for the energy (resp. entropy) for $\epsilon=0.1$ and $\epsilon=10^{-5}$. We interpret these differences as an estimation of the finite time step error at $\epsilon$. (The results for the energy with $\epsilon=0.1$ and $\epsilon=0.01$ are already identical up to 8 decimals. This is why we have not shown them.) We observe from these tables that the error on the entropy is about ten times larger than that on the energy and that both increase around the pseudo critical point $\lambda_\chi^*$. Simulations with $\chi=4$ and $\chi=8$ show similar results. If we now compare the values of the energy  and entropy yielded by our simulations with the exact values for an infinite chain (Table \ref{tab:error-exact} and Table 
\ref{tab:error-exact-entr}), we see that the errors are larger in the vicinity of the critical point and that, as expected, they decrease as we decrease $\epsilon$.

\begin{table}[ht]
\begin{tabular}{ c c c c c c }
$\lambda$ & Exact & $\chi = 2$ & $\chi=4$ &  $\chi=8$ & $\chi=16$ \\
\hline\hline
0.5 &   1.06354440 &  33 &     $< 0.1$ &   $< 0.1$  &  $< 0.1$ \\
0.6 &   1.09223858 &  172 &    $< 0.1$  &   $< 0.1$  & $ < 0.1$\\
0.7 &   1.12682867 &  745 &    2  &  $ < 0.1$  & $ < 0.1$\\
0.8 &   1.16780951 &  2978  &   12  &  $ < 0.1$  &  $< 0.1$ \\
0.9 &   1.21600091 &  12173  &  126  &   $< 0.1$  & $ < 0.1$\\
1.0 &   1.27323954 &  69712  & 4683  & 261  & 15 \\
1.1 &  1.34286402  &  146576  & 1642  &   6  & $ < 0.1$ \\
1.2 &   1.41961927 &  77696  &  416 &    $< 0.1$  &  $< 0.1$ \\
1.3 &   1.50082324 &  45675  &  141 &  $ < 0.1$ &  $< 0.1$ \\
1.4 &   1.58518830 &  28719  &   57  &   $< 0.1$  &  $< 0.1$ \\
1.5 &   1.67192622 &  18964  &   25  &   $< 0.1$  &  $< 0.1$ \\
\hline\hline
\end{tabular}
\caption{Errors in the energy in relation to the exact value for $\chi=2, 4, 8 \mbox{ and } 16$. These
errors are greater around the critical point (which for $\chi=2$ is close to $\lambda=1.1$). These
values were obtained with $\varepsilon=0.1$, showing a clear dominance of truncation error over the errors introduced by finite step time evolution. All
entries of this table should be multiplied by $10^{-8}$.}\label{tab:error-exact} 
\end{table}

\begin{table}[ht]
\begin{tabular}{c c c c c c}
$\lambda$ & Exact & $\chi = 2$ & $\chi=4$ &  $\chi=8$ & $\chi=16$ \\
\hline\hline
0.5 & 421292 & 	-3	&	-3	&	$<0.1$	 &	$<0.1$ \\
0.6 & 914778  &	-36	 &	-36	 &	$<0.1$	 &	$<0.1$\\
0.7 & 1869961	&	-389	 &	-389	 &	$<0.1$	 &	$<0.1$\\
0.8 & 3804448 &	-4255	 &	-4255	 &     -1 & $<0.1$ \\
0.9 & 8484551 &	-66920	 &	-66920	 &	-12 & -2 \\
1.1 & 47444179 &	-437545	 &	-437545	 &	-6473 & -3 \\
1.2 & 36551466 &	-74387	 &	-74387	 &	-270 &  5 \\
1.3 & 30064632 &	-20221	 &	-20221	 &	-23  & 5 \\
1.4 & 25539496 &	-6976	 &	-6976	 &	$<0.1$   &5\\
1.5 & 22144107 &	-2797	 &	-2797	 &	4 & 5 \\
\hline\hline
\end{tabular}
\caption{Errors in the entropy  for different
values non-critical $\lambda$ and of $\chi$. All values have
been multiplied by $10^8$. Note the increasing accuracy as a function of $\chi$.}
\label{tab:error-exact-entr}
\end{table}

\begin{figure}[hbt]
\includegraphics[width=8cm,height=5cm]{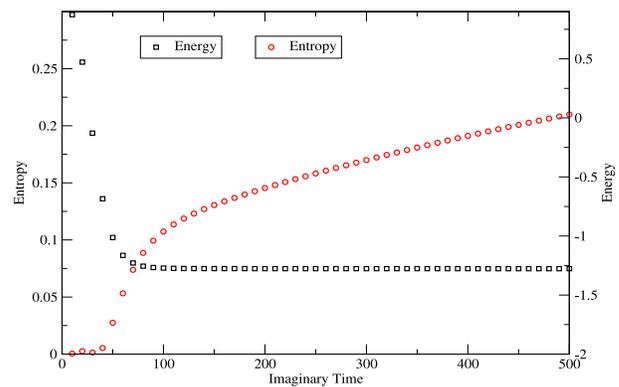}
\caption{Convergence of the energy and entropy,  at the effective critical point, during the imaginary time evolution, with $\chi=8$, $\lambda=1.006$ and $\varepsilon=10^{-2}$. The full convergence of the energy (eight decimals) took $\sim 10^5$ steps while $\sim 6\ 10^5$ steps where necessary to make the entropy converge.}\label{fig:convergencetime}
\end{figure}

Let us now clarify why we were interested in reaching full convergence for the half-chain entropy. A common method to locate a phase transition is to analyze the variation of an order parameter. On another hand, we know that the half-chain entropy of a critical system diverges while, off and close to criticality, it scales as the logarithm of the correlation length and thus remains finite \cite{lato04}. It is therefore reasonable to think of using the half-chain von Neumann entropy, to detect a phase transition. It turns out that when running the \texttt{iTEBD} algorithm, $S$ converges faster to a steady value than the mean value of the order parameter and thus provides a faster detection of  the position of the critical point (varying the magnetic field, $\lambda$, and scanning for the peak of $S$). Yet, the von Neumann entropy converges more slowly than the energy, see Fig. \ref{fig:convergencetime}. Around the critical point, the spectrum of the Hamiltonian is filled with a lot of low-energy excited levels which energy is very close to that of the ground state. An arbitrary superposition of such excited states will have energy close to that of the ground state, but can in principle exhibit very different entanglement properties.  We believe that this is why it takes  much longer to get a reliable estimate of the entropy. One has to make the energy converge close enough to that of the ground state so that the entropy of the obtained state also faithfully reflects that of the ground state. 

\section{Metastabilities}
%%%%%%%%%%%%

\begin{figure}[hbt]
\includegraphics[scale=0.34]{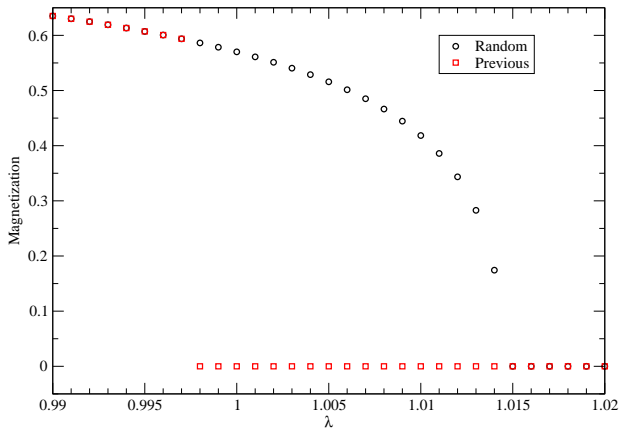}
\vspace{0.1cm}
\caption{Magnetization as a function of the transverse magnetic field using different initial states for
$\chi=4$ and $\varepsilon=0.1$. In one case (open circles) we use a random $2 \times 2$ matrix 
as an initial state. In the second case (open squares) the initial state for $\lambda=\lambda_0$ is the final state obtained for $\lambda=\lambda_0+0.001$. The two methods do not give similar results for 
the position of the critical point.}\label{fig:histerese} 
\end{figure}

An important issue when running the \texttt{iTEBD} algorithm is to be sure that one is not driven to a local minimum. Here we point out the existence of some  meta stabilities in the simulation with respect to the choice of the initial state (an effect which is also
present in standard DMRG simulations). In all our calculations, we have used an initial state which matrices $\Gamma_A$ and $\Gamma_B$ (see \cite{Vidal:infinite-chain} for details) are of the following form : random entries in the $2 \times 2$ left upper corner, and all other entries set to zero. However, when performing a simulation for the Ising chain for some value $\lambda$ of the transverse field, one could use,  as initial state, the result of a simulation performed at some
close value of $\lambda$. Although this procedure can substantially decrease the time necessary to make the energy and the half-chain entropy converge, it can also lead to misleading results regarding the position of $\lambda^*_{\chi}$, taken as the point where the order parameter vanishes,  as can be seen on Fig. \ref{fig:histerese}. Simulations which start from a previous minimization
run of a larger $\lambda$ do produce unphysical results. Thus, all simulations must start from random
initial conditions. 

\vskip .4cm

\section{Boosted \texttt{iTEBD}}
%%%%%%%%%%%%%%%

The performance of the \texttt{iTEBD} algorithm depends  on the initial
conditions and the gap above the ground state. The results of our
study  suggest  that using finite $\chi$ one is perturbing the
system in a way similar to have an effective gapped Hamiltonian.
However, if the gap is small the convergence of the algorithm can be very slow. To see
this, we can consider
as initial state a state with non zero projection on the ground state
\begin{equation}
  | \psi \rangle = \alpha| \psi_0 \rangle + \sqrt{(1 - \alpha^2)} |
\psi_{\perp} \rangle
\end{equation}
with $\vert\alpha\vert<1$.
It is easy to see that,
if the Hamiltonian has a gap $\Delta$, the Euclidean  evolution of an
initial state with non zero projection on the ground state will lead to:
\begin{equation}
  | \psi' \rangle = \exp (- H \tau) | \psi \rangle = \alpha \exp (-
E_0 \tau) | \psi_0
  \rangle + \sqrt{(1 - \alpha^2)} | \psi'_{\perp} \rangle
\end{equation}
with $| \psi'_{\perp} \rangle =\exp (- Ht) | \psi_{\perp} \rangle $.
From
\begin{equation}
  \langle \psi'_{\perp} |H| \psi'_{\perp} \rangle \geqslant \Delta+E_0,
\end{equation}
we see the long time limit of the above expression, differs from
the ground state (as already pointed out in ref. \cite{Vidal:first})
by terms of the order:
\begin{equation}\label{eq:overlap}
  | \langle \psi_0 | \psi' \rangle | \sim 1 - \frac{1 - \alpha^2}{2
\alpha^2} \exp (- 2\tau   \Delta ) .
\end{equation}
Now if we approach the critical point of a phase transition we know
that the correlation length scales
 with the critical index $\nu$ of the corresponding universality
class
\begin{equation}
  \xi \sim t^{-\nu},
\end{equation}
where $t$ denotes, again, the distance from the critical point.
Assuming that $ \Delta\sim 1 /\xi$, we see that, even  in the case of
a good guess of the initial state (that is, in the case
$\vert\alpha\vert^2 \sim 1$), the convergence of the algorithm slows
down in the critical region. In order to partially cure this slowing down
we can perform a linear extrapolation  of the results obtained
after a small interval of Euclidean time $d \tau$ and get a new
estimate for the MPS.
Given a generic  element of the MPS matrix at Euclidean time
$\tau$, $A(s_i)(\tau)$, and the same element at time $\tau+d\tau$,
$A(s_i)(\tau+d\tau)$,
we construct a new MPS which matrix elements $A(s_i)(T)$ are the
extrapolation  at time $T$ of the straight line passing from the two
points at $\tau$ and $\tau+d \tau$.
Before promoting the guessed MPS to a new initial condition, we should
check that the state it describes has a lower energy than the MPS
obtained at $\tau+d \tau$ before the extrapolation.
A lower energy indeed means a greater overlap with the ground state.
In this case the extrapolation is successful and we promote the guess
to an initial condition
of the new evolution.
In case the energy of the new extrapolated  state is greater than the
energy of the state before the extrapolation, we just neglect it and
keep the state we had before the extrapolation.  The new Euclidean
evolution is also of  length $d\tau$ and is followed by the  attempt
of a new extrapolation. We iterate the procedure till we reach
convergence. We call this technique \texttt{ boosted iTEBD}.

\begin{figure}[hbt]
\includegraphics[height=5cm,width=7.2cm]{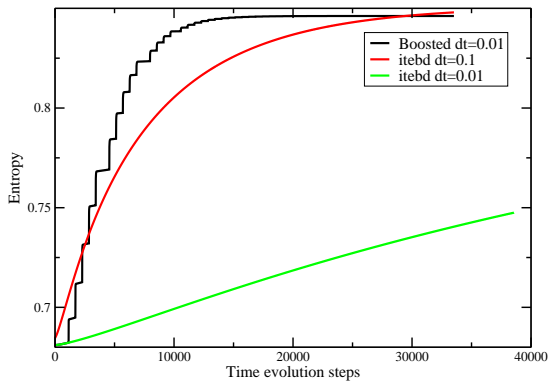}
\caption{ {\texttt Boosted itedb} algorithm. We plot the half-chain
entropy as a function of the Trotter steps for a $\chi=32$ MPS at
$\lambda=1$. This is taken as a typical case from a large number of
examples with different $\chi$ and magnetic fields that present a
similar behavior. We compare the results obtained with
$\varepsilon=0.01$ with both boosted and standard \texttt{iTEBD} and
the results obtained with $\varepsilon=0.1$ with standard
\texttt{iTEBD}. As we can see,  the unboosted case with
$\varepsilon=0.01$ is far from having  converged in the number of
Trotter steps considered. On the other hand, the boosted system with
$\varepsilon=0.01$ converges (up to 8 decimals) in a smaller amount of
time steps than the unboosted algorithm with an $\varepsilon$ ten
times bigger. Indeed, the latter simulation has still not converged in
the window of Trotter steps shown. The discrepancy in the asymptotic
values is due to $ \varepsilon$ corrections described in the previous
appendices. From this plot, we can safely deduce that the effect of
the boost is to reduce the convergence time by a factor greater than
$10$ in the case we have analyzed.} 
\label{fig:boost}
\end{figure}

In order for the extrapolation to work, we have to tune finely its two
parameters: the waiting time $d \tau$ and the amount of time we
extrapolate, $T$. Once these two parameters are fixed, we are able to
accelerate the convergence by a factor greater than $10$. A typical
case is shown in Fig. \ref{fig:boost} where we show 
the convergence of the simulations of a $\chi=32$ MPS at
$\lambda=1$. Without the   \texttt{boosted iTEBD} algorithm, with
$\varepsilon=0.01$, after the number of Trotter steps considered, the
system had still not converged. Increasing  $\varepsilon=0.1$
translates in a coarser precision but with a convergence time about
ten times shorter. A further improvement in convergence is obtained by
keeping the same $\varepsilon=0.01$, and hence the same precision, but
boosting the evolution with the extrapolation technique described. We
see that the gain in convergence time is bigger  by a factor of ten.
As we can see, there is a point where the extrapolation fails and the
normal evolution is continued. We can also check that before the first
extrapolation, the boosted evolution coincides with the unboosted
evolution with the same $\varepsilon$.

\vskip 1cm

\section{Comparison with DMRG}

In order to ensure that the effects we observe are not artifacts of the algorithm used, we reproduced some of them with a different algorithm.
We have chosen to use the open source code for DMRG written by the Pisa group \footnote{This part of the work has been developed by using the DMRG code released within the "Powder with Power" project (www.qti.sns.it)}.  This program performs an infinite DMRG update of the system by growing it till it reaches a chosen chain length. At this stage, it performs several finite size sweeps through the chain (at least three in our case) in order to compute the reduced density matrix of all possible chain bipartitions and improve the infinite results \cite{DMRGdummy} . 

\begin{table}[ht]
\begin{tabular}{c c c }
&Energy &Entropy \\
\hline\hline
DMRG N=16284&			-1.27321717&	0.68557374\\
iTEBD&			        -1.27323939&	0.68065196\\
iTEBD - DMRG(N=16384)&		-0.00002222&   -0.00492178\\
\hline\hline
\end{tabular}
\caption{Comparison of DMRG energy and entropy results  with the infinite
size result produced by {\texttt iTEBD} with $\epsilon=10^{-4}$ and where both methods used $\chi=16$. }
\label{tab:dmrg-itebd}
\end{table}

We checked the stability of the presented results on the variation of the number of finite size sweeps. In this way, we are sure that the results have converged. We have checked that for a fixed number of level ($m$ in the language of DMRG that corresponds to $\chi$ in this paper), increasing the chain length makes the results converge to those obtained with the algorithm we have used in the paper. It was interesting to see that  the DMRG convergence is however quite slow as compared with the \texttt{boosted iTEBD}.

\end{document}